\documentclass[twocolumn]{aastex62}

\graphicspath{{./}{figures/}}

\usepackage{graphicx}
\usepackage{amsmath}
\usepackage{amssymb}
\usepackage{gensymb}
\usepackage{lineno}
\usepackage{float}

\submitjournal{ApJ}

\shorttitle{DYNAMICAL FORMATION OF AN INTERMEDIATE MASS BLACK HOLE}
\shortauthors{Anagnostou et al.}

\graphicspath{{./}{figures/}}

\begin{document}

\title{REPEATED MERGERS OF BLACK HOLE BINARIES: IMPLICATIONS FOR GW190521} 

\correspondingauthor{Oliver Anagnostou}
\email{oanagnostou@student.unimelb.edu.au}
\affiliation{School of Physics, The University of Melbourne, VIC 3010, Australia}

\author{Oliver Anagnostou}
\affiliation{School of Physics, The University of Melbourne, VIC 3010, Australia}

\author{Michele Trenti}
\affiliation{School of Physics, The University of Melbourne, VIC 3010, Australia}
\affiliation{Australian Research Council Centre of Excellence for Gravitational Wave Discovery (OzGrav)}
\affiliation{Australian Research Council Centre of Excellence in All Sky Astrophysics in 3 Dimensions (ASTRO 3D)}

\author{Andrew Melatos}
\affiliation{School of Physics, The University of Melbourne, VIC 3010, Australia}
\affiliation{Australian Research Council Centre of Excellence for Gravitational Wave Discovery (OzGrav)}

\begin{abstract}
The gravitational wave event GW190521 involves the merger of two black holes of $\sim 85\text{M}_\odot$ and $\sim 66\text{M}_\odot$ forming an intermediate-mass black hole (IMBH) of mass $\sim 142\text{M}_\odot$. Both progenitors are challenging to explain within standard stellar evolution as they are within the upper black-hole mass gap. We propose a dynamical formation pathway for this IMBH based on multiple mergers in the core of a globular cluster. We identify such scenarios from analysis of a set of 58 N-body simulations using \texttt{NBODY6-gpu}. In one of our simulations, we observe a stellar black hole undergoing a chain of seven binary mergers within 6 Gyr, attaining a final mass of  $97.8\text{M}_\odot$. We discuss the dynamical interactions that lead to the final IMBH product, as well as the evolution of the black hole population in that simulation. We explore statistically the effects of gravitational recoil on the viability of such hierarchical mergers. From the analysis of all 58 simulations we observe additional smaller chains, tentatively inferring that an IMBH formation through hierarchical mergers is expected in the lifetime of a median mass globular cluster with probability $0.01 \lesssim p \lesssim 0.1$ without gravitational merger recoil. Using this order-of-magnitude estimate we show our results are broadly consistent with the rate implied by GW190521, assuming that gravitational recoil ejection of progenitors has a low probability. We discuss implications for future gravitational wave detections, emphasising the importance of studying such formation pathways for BHs within the upper mass gap as a means to constrain such modelling.

\end{abstract}


\section{Introduction} 

\label{sec:intro} 
On 2020 September 2\citep{2020PhRvD.102d3015A}, the Laser Interferometer Gravitational-Wave Observatory (LIGO) and Virgo collaboration announced the detection of a gravitational wave signal consistent with a Black Hole Binary (BHB) merger producing a remnant BH of mass $142\pm^{28}_{16}\text{M}_\odot$ \citep{2020PhRvL.125j1102A}. The event, known as GW190521, involved the merger of component masses $m_1 = 85\pm^{21}_{14}\text{M}_\odot$ and $m_2 = 66\pm^{17}_{18}\text{M}_\odot$. These are the heaviest two merging BHs yet observed by LIGO/Virgo \citep{2016PhRvL.116f1102A,PhysRevX.6.041015,PhysRevLett.116.241103,PhysRevLett.118.221101,Abbott_2017,2019PhRvX...9c1040A,stoyan_binnewies_friedrich_2008, 2020ApJ...896L..44A, 2020PhRvD.102d3015A, 2020ApJ...892L...3A}. A similarly heavy merger was announced in the second half of LIGO's third observational run, know as GW200220\_061928 \citep{theligoscientificcollaboration2021gwtc3}. Interestingly, both events have component masses that potentially lie within the supposed upper black hole mass gap, a range of BH masses which is outside what is predicted through standard stellar evolution. The mass gap is thought to originate from pair-instability supernovae \citep{Marchant_2019, 2003ApJ...591..288H, Belczynski_2016, Spera_2017, 2017ApJ...836..244W, 2019ApJ...878...49W}. Current theoretical modelling places the lower cutoff at $\sim {50\text{M}} ^{+4}_ {-8}\text{M}_\odot$ \citep{Marchant_2019, Farmer_2019, 2020ApJ...902L..36F, Marchant_2020, Renzo_2020, Leung_2019, Farmer_2019, Belczynski_2020, Mapelli_2020, Costa_2020}, although there are uncertainties pertaining to hydrodynamics, relativistic effects and nuclear physics. The modelling therefore implies that at least the primary masses are well within the mass gap and \citep{2016PhRvL.116f1102A} cannot have originated directly from a stellar progenitor. 

The existence of physics beyond the standard model could lead to modifications in the canonical pair-instability supernova process, enabling the production of heavier stellar BHs than current theories permit. Phenomena such as the existence of new light particles \citep{2021PDU....3200801C}, a different magnetic dipole moment of the electron neutrino \citep{Barger_1999,Heger_2009} and the addition of extra spatial dimensions \citep{padilla2015lectures} can all alter the supernova process to produce heavier BHs. Indeed \citet{2020PhRvL.125z1105S} propose such new physics as potential explanations for GW190521.  

\medskip \noindent
However, without invoking non-standard physics, there are other possible mechanisms that can form higher mass BHs, particularly in dense environments through dynamical interactions. If a Globular Cluster (GC) is sufficiently dense at formation, massive stars can undergo one or more collisions before stellar core collapse. The result may be an evolved star with an oversized hydrogen envelope, which collapses to form a BH within the upper mass gap \citep{Spera_2019}. \citet{Kremer_2020} explored the formation of BHs with masses within the  upper mass gap through collisions of young massive stars in dense star clusters. They found that $\sim20\%$ of BH progenitors undergo one or more collisions before collapse, leading to $\sim1\%$ of all stellar BHs forming within the upper mass gap. \citet{Di_Carlo_2020} found a slightly higher fraction of mass gap BHs through young stellar collisions, with up to $\sim 6\%$ of all simulated BHs $\gtrsim 60 \mathrm{M}_{\odot}$, depending on the progenitor’s metallicity.

\medskip \noindent
BHBs may also form and evolve via gravitational interactions with other compact bodies in a dense stellar environment such as a GC \citep{2006ApJ...637..937O,2010MNRAS.402..371B,2007PhRvD..76f1504O,2013MNRAS.435.1358T,2018MNRAS.480.2343C}. In these dense environments, the BH product of a previous BHB merger can go on to form a new BHB and merge again, which we henceforth refer to as a second-generation BH. This process can continue, creating a chain of mergers \citep{O_Leary_2016, Fishbach_2017, PhysRevD.95.124046, PhysRevLett.123.181101, Antonini_2019, PhysRevD.95.124046, Samsing_2018, PhysRevD.100.043027, Gerosa_2020, Safarzadeh_2019, Gayathri_2020, 2020ApJ...900..177K, Doctor_2020, Baibhav_2020}. Such hierarchical mergers allow BHs to reach masses above what is achievable through stellar evolution. In particular, it has been shown that GCs are effective nurseries to host such hierarchical mergers, and that populating the mass gap is possible through this dynamical channel \citep{PhysRevD.95.124046, 2021ApJ...923..126S}.

\medskip \noindent
While intermediate mass black holes (IMBHs) with $M_{\text{IMBH}}\gtrsim 100~\mathrm{M_{\odot}}$ have not been reported to originate in dynamical simulations of hierarchical BHB mergers, indicating such a scenario is uncommon, recent work suggests that repeated mergers between BHs and very massive stars in young star clusters are a promising channel to form an IMBH within the first $15$ Myr after star cluster formation \citep{2021MNRAS.501.5257R}. One of the main barriers to forming an IMBH through hierarchical mergers is the potentially large gravitational recoil velocities of merger products \citep{PhysRev.128.2471,1973ApJ...183..657B,1984MNRAS.211..933F, 2010ApJ...719.1427V, 2007ApJ...659L...5C, Lousto_2010}. Such recoil velocities can exceed a typical cluster's escape velocity, ejecting the second-generation BH and preventing future mergers. Although full calculations for gravitational recoil require numerical relativity, researchers have derived analytic formulae that can reasonably approximate the recoil behaviour seen in numerical simulations \citep{Lousto_2010, 2010ApJ...719.1427V}.

\medskip \noindent
Recently, various research groups investigated repeated BH mergers within star clusters with the inclusion of gravitational merger recoil, either through Monte Carlo \citep{Rodriguez_2019} or direct N-body methods \citep{2020MNRAS.tmp.2087B,https://doi.org/10.48550/arxiv.2109.14612}. In our work, gravitational recoil is not calculated while the simulations run, but it is instead implemented via post-processing. This choice offers a hybrid solution that affords benefits of a Monte Carlo approach while having a direct N-body code for highly accurate stellar dynamics. In particular, we are able to turn recoil kicks on and off to explicitly investigate the effects of recoil on BHB merger chains. We are also able to investigate the distribution of possible recoil velocities by sampling over different BH spin orientations for a given merger, letting us explore the probability of finding different length merger chains within GCs. The latter would be particularly challenging to investigate otherwise because of the limited number of BH mergers in a single simulation and the computational cost of running direct N-body simulations.


\medskip \noindent
In this paper, we use a set of direct N-body simulations of dense star clusters that include both gravity and stellar evolution \citep{2020PASA...37...44A} to investigate hierarchical BHB mergers as a potential formation channel for IMBHs within GCs over timescales compared to their typical ages (i.e. $\sim 10$ Gyr). In Section \ref{sec:method}, we briefly describe our simulation framework, including initial conditions and our treatment of gravitational recoil. In Section \ref{sec:snowball}, we explore in detail one simulation that shows the evolution of a specific BH --- henceforth referred to as ``Snowball" ---  undergoing seven BHB mergers in 6.5 Gyr to reach a final mass $\text{m}_{\text{BH}}=97.8 M_{\odot}$. In Section \ref{sec:The Merger Chain With Recoils}, we discuss the effects of gravitational recoil on the viability of such a merger chain, highlighting limitations in the calculations. From the length distribution of merger chains within the full set of simulated clusters, in Section \ref{sec:rates} we estimate the comoving rate of IMBH formation in a typical GC. Finally, in Section \ref{sec:implications} we discuss implications of this formation channel for the GW190521 detection, as well as for future gravitational wave detections. 

\section{Method}
\label{sec:method}

\subsection{Simulations}
\label{subsec:simulations}

This work is based on a set of 58 direct N-body simulations of mid-sized star clusters previously presented in \cite{a2020PASA...37...44A}, run using \texttt{NBODY6} with GPU support \citep{2012MNRAS.424..545N}. The set of simulations contains $5\times 10^4 \leqslant N \leqslant 2\times10^5$ initial stars, and utilizes a range of initial conditions (see \citealt{2020PASA...37...44A} for more detail). 

\medskip \noindent
Single stars and binaries evolve according to the ``SSE" and ``BSE" algorithms respectively \citep{2000MNRAS.315..543H,2013ascl.soft03014H}, which are designed to handle complex evolutionary processes, including common envelope evolution and mass transfer, collisions and supernova kicks. The version of SSE available in our simulation code utilizes the \citet{2000MNRAS.315..543H} supernovae scheme, which is based on work done by \citet{1989ApJ...347..998E}. Under these prescriptions SSE calculates BH birth masses based on the the formulation given by \citet{2008ApJS..174..223B}, whereby BH mass is soley a function of progenitor star core mass. Note that the code version we use does not include any prescriptions for oversized black holes, such as those presented in \citet{Spera_2019}. As stars evolve they lose mass and angular momentum due to stellar winds. These winds are calculated using the semi-empirical prescriptions of \citet{2010ApJ...714.1217B} whereby mass loss rate is a function of metallicity, stellar surface temperature and luminosity. One of the key features of this model is a "bi-stability jump" that occurs at surface temperatures of $\sim 25000$ K, which leads to a significant increase in mass loss due to an increase in the ratio of Fe III to Fe IV, which is a more effective wind driver. Binary orbital changes due to gravitational waves and subsequent merger times are calculated by integrating the orbit-averaged Peters equations \citep{PhysRev.136.B1224}, \texttt{NBODY6} does not include post-Newtonian corrections to the equations of motion.

\medskip \noindent
In this paper we focus on a specific simulation, with the following specifications: initial particle number $\text{N}_0 = 100,000$ particles, primordial binary fraction $f = 0$, metallicity $Z$ = 0.001, initial half mass radius $r_\mathrm{h,0} = 2.5$ pc and dimensionless King concentration parameter $W_0 = 7$. Stellar masses are drawn from a \citet{2001MNRAS.322..231K} initial mass function, between 0.08 $\text{M}_\odot$ and 100 $\text{M}_\odot$. The initial core radius is 0.56 pc with a tidal radius of 71.5 pc. The cluster initially under-fills its tidal radius by a factor of three. The tidal radius is calculated by placing the cluster into a circular orbit around a point-mass galaxy with a galactocentric distance of $23.3$ $\text{kpc}$. We also initialize the cluster in dynamical equilibrium and with no mass segregation. The cluster model is representative of an average mid-sized GC (see \citealt{2003gmbp.book.....H} Table 1.1). The stellar metallicity of Z = 0.001 ([Fe/H] $\approx -1.33$) is approximately the median metallicity for Milky Way GCs \citep{1996AJ....112.1487H}.The cluster is evolved for a total of 12.5 Gyr. 

\medskip \noindent
Natal kicks are applied to neutron stars and black holes (no natal kicks are given to white dwarfs), drawn from a Maxwellian distribution with velocity dispersion $\sigma_k \approx 1000$ ms$^{-1}$, the initial velocity dispersion of the cluster (see \citealt{2019MNRAS.485.5752D} for further explanation). Our choice of Maxwellian velocity dispersion is significantly smaller than the default \texttt{NBODY6} value of $\sigma_k \approx 265$ kms$^{-1}$, which is based on pulsar proper motion observations \citep{2005MNRAS.360..974H}. We choose to impart low natal kicks to create conditions such that a large fraction of BHs are retained in the model clusters after formation. This is justified as it somewhat mimics the effects of fallback modulated BH natal kicks on BH retention statistics \citep{2013MNRAS.434.1355J,2015MNRAS.453.3341R,2020A&A...639A..41B}. This natal kick treatment is the same for neutron stars and BHs, as the public version of \texttt{NBODY6} does not contain prescriptions for mass fallback or momentum conserving BH natal kicks. One caveat of these low natal kicks is that an unusually large number of neutron stars are retained inside the host cluster after formation compared to similar studies \citep{2020ApJ...888L..10Y, 2022ApJ...934L...1K,2020ApJS..247...48K, 2017ApJ...834...68C}. However, for most of a clusters lifetime neutron stars are too light to be part of the set of most massive particles that migrate to the cluster core. Hence the presence of neutron stars is unlikely to significantly effect the dynamics of the cluster core until very late in the cluster's evolution. In addition, in the current study we are only interested in BHB interactions and mergers, and so the effects of the high neutron star retention fraction on our results are expected to be minimal. For information on the complete set of simulations, further detail on the code used, and full discussion of initial conditions see \citealt{2020PASA...37...44A} and \citet{2019MNRAS.485.5752D} (see also \citealt{2016ApJ...819...70M}). 
 
\subsection{Gravitational Recoil}
\label{subsec:Gravitational Recoil} 

Recent investigations implemented gravitational recoil calculations on-the-fly  \citep{2018PhRvL.120o1101R, 2018MNRAS.481.2168M,2020MNRAS.tmp.2087B}. The public version of \texttt{NBODY6} does not include gravitational recoil nor the effects of BH spin, thus, we added these additional physical ingredients in post-processing. We calculate kick velocities using a numerical relativity based fitting formula (detailed below) in a similar manner to \citet{2018PhRvL.120o1101R}, \citet{2018MNRAS.481.2168M} and \citet{2020MNRAS.tmp.2087B}. 

\medskip \noindent

The underlying spin magnitude distribution for stellar BHs is currently unknown and not well constrained. Two previous studies have simply assumed a uniform spin magnitude distribution in which spins are randomly assigned at birth \citep{2018PhRvL.120o1101R,2018MNRAS.481.2168M}. Unlike those studies, we assign spin magnitudes using a fitting formula based on the \texttt{MESA} stellar evolution code, leading to a non-uniform spin magnitude distribution (see equation 4 in \citet{Belczynski_2020} for fitting formula). \texttt{MESA} calculates spins by modelling the loss of progenitor core angular momentum via stellar winds \citep{Paxton_2010}. The recoil velocities are then calculated using the analytic approximation \citep{2010ApJ...719.1427V} 


\begin{equation} \label{eq:1}
    \textbf{V} = V_{\bot m} \hat{e}_1 + V_{\bot s} \left(\cos(\xi)\hat{e}_1 + \sin(\xi)\hat{e}_2\right) + V_\parallel \hat{e}_3,
\end{equation}

\noindent where the components perpendicular to the orbital axis are defined as

\begin{equation}
\label{eq:2}
    V_{\bot m} = A \eta^2 \sqrt{1 - 4\eta}\left(1 + B \eta \right),
\end{equation}

\noindent and 

\begin{equation}
\label{eq:3}
    V_{\bot s} = H \frac{\eta^2}{\left(1 + q\right)}\left[|\alpha^\parallel_2| - q |\alpha^\parallel_1| \right],
\end{equation}

\noindent
and the component parallel to the orbital axis is

\begin{gather} \label{eq:4}
    V_\parallel = \nonumber \\ 
    \frac{K_2 \eta^2 + K_3 \eta^3}{q + 1} \left[q|\alpha^\bot_1| \cos(\phi_1 - {\Phi_1}) - |\alpha^\bot_2| \cos(\phi_2 - {\Phi_2})\right] \nonumber \\ 
    + \frac{K_S\left(q - 1\right)\eta^2}{\left(q+1\right)^3} \left[q^2|\alpha^\bot_1| \cos(\phi_1 - {\Phi_1}) - |\alpha^\bot_2| \cos(\phi_2 - {\Phi_2})\right]. 
\end{gather}

\noindent
Here $A$, $K_2$, $K_3$, and $K_S$ are constants (units: $\text{kms}^{-1}$), $B$ is a dimensionless constant, and $\xi$ measures the angle between the unequal mass and spin contribution to the recoil velocity in the orbital plane, and is taken to be 145\degree \citep{Lousto_2010}. The mass ratio is defined as $q = m_2/m_1 \leqslant 1$, and the symmetric mass ratio as $\eta = q/(1+q)^2$, where $m_1$ and $m_2$ are the primary and secondary black hole masses respectively. Additionally, $\hat{e}_1$ is the unit vector in the direction of the separation between the two BHs at the end of the last orbit prior to the final plunge, $\hat{e}_3$ is the unit vector in the direction of the orbital axis, and $\hat{e}_2 \equiv \hat{e}_1 \times \hat{e}_3$. The symbols $\perp$ and $\parallel$ refer to the components of the dimensionless spin vectors perpendicular and parallel to the orbital angular momentum respectively. For example, $\alpha^\bot_i$ represents the projection of the i'th black hole’s dimensionless spin onto the orbital plane, where $i = 1$ or 2. The symbol $\phi_i$ represents the angle between $\alpha^\bot_i$ and the displacement vector between the black holes, as measured at the end of the last orbit before the final plunge. The symbol ${\Phi_i}$ represents the amount by which this angle precesses before the merger. The ${\Phi_i}$ parameters are not constants as they depend on the specifics of a BHB inspiral, such as the mass ratio, initial separation and spins. For the purposes of the current work, we set ${\Phi_1}$ = ${\Phi_2} = 0$, as suggested by \citet{Lousto_2010} for statistical simulations of BHB mergers. For more details on these analytic formulae, including parameter fitting, see \citet{2010ApJ...719.1427V}. 

\medskip \noindent

Spin directions are expected to be randomly oriented for dynamically formed BHBs. The relative orientation of two BH spin vectors can have a significant impact on the recoil velocity. To deal with this large range of potential kick velocities based on spin orientation, we adopt a Monte Carlo method. For all mergers found by \texttt{NBODY6}, we randomly assign a spin direction to the merging BHs, drawing from an isotropic distribution. All other properties such as mass, initial separation and spin magnitudes are fixed, as taken from \texttt{NBODY6} and \texttt{MESA}. We calculate the corresponding merger recoil velocity for the second-generation BH. If the resulting BH velocity (vector sum of the binary's centre of mass velocity prior to merger and the recoil velocity) exceeds the cluster escape velocity at the time of merger, then we exclude it from any future mergers, treating the BH as if it escaped the cluster. For any post-merger BH that is retained after recoil, we assign a new spin magnitude via numerical relativity based fitting formulae presented in \citet{2008PhRvD..78d4002R} (eq 8,10). We repeat this post-processing procedure $10^4$ times for the full merger history generated by each of the 58 simulations, in order to compute the probability of various escape and retention outcomes attached to every merger chain in every simulation by assuming randomized spin. Because of the variable merger recoil velocities between each calculation, the number of mergers is not always the same, as some hierarchical mergers can be cut short due to the ejection of one of the post-merger BHs. 

\medskip \noindent
This post-processing implementation of recoil does come with some inherent disadvantages. Since we are not applying recoil velocities directly in \texttt{NBODY6}, our method only approximates the full effects of recoil kicks. Any post-merger BH that would be ejected due to recoil still remains in the cluster (as merger recoils are only calculated in post-processing), we simply ignore any future mergers involving it. Keeping these massive BHs in the cluster when they should be ejected does impact the dynamics of the cluster, particularly in the dense core. It may lower the probability of other BHB mergers, as BHs tend to primarily interact with the most massive BH in the core. However, this method does account for the most important impact of kicks when regarding hierarchical BHB mergers --- the prevention of long merger chains due to the ejection of merger products. 

\medskip \noindent
However, there are distinct advantages to calculating recoil outside of the N-body code. With this approach, we can enable and disable recoil kicks for our set of simulations - meaning we can directly assess the impact of merger recoils on the distribution of merger chain lengths. Furthermore, our Monte Carlo method for assigning spin magnitudes inherently requires post-processing. In fact, when recoils are calculated within
the simulation code, a single spin orientation must be selected, leading to a specific recoil velocity
and specific outcome for the post-merger BH. Given the rarity of BH mergers in a single simulation and the computational resources required to run a single realization of a direct N-body simulation, it would be practically impossible to build sufficient information on the expected distribution of BH recoil events from run-time recoil implementation.

\section{The Merger Chain Without Recoils}
\label{sec:snowball}

We start by studying the merger chain without merger recoils caused by gravitational radiation reaction. This study serves two goals. (i) It provides a control experiment, or baseline, against which the strength and impact of the recoil effect can be assessed and quantified. Having such a benchmark allows us to directly investigate the impact of recoils in isolation by turning them on and off. This is important at a time when general relativistic fitting formulas for recoil outcomes are still under development and not yet settled. (ii) It yields an estimate of the maximum plausible mass of an IMBH like Snowball within our cluster models. This upper bound is important, because the physics of merger recoils remains a topic of active research in numerical relativity, and the recoil prescriptions predicted by theory are likely to be refined in the future. Note that Snowball's final mass does not represent the true upper bound for the mass of a BH through repeated merger, merely it is an estimate for the upper bound in our cluster models.

\medskip \noindent 
In addition, the details of how a long merger chain occurs without recoil can still provide insight into chains with recoil, as long merger chains, although less likely, are still possible with gravitational recoil and should evolve via the same dynamical processes, as demonstrated explicitly in this paper. Understanding these dynamical processes vital to progressing our understanding of repeated mergers.

\begin{figure}
    \centering
    \includegraphics[width = \columnwidth]{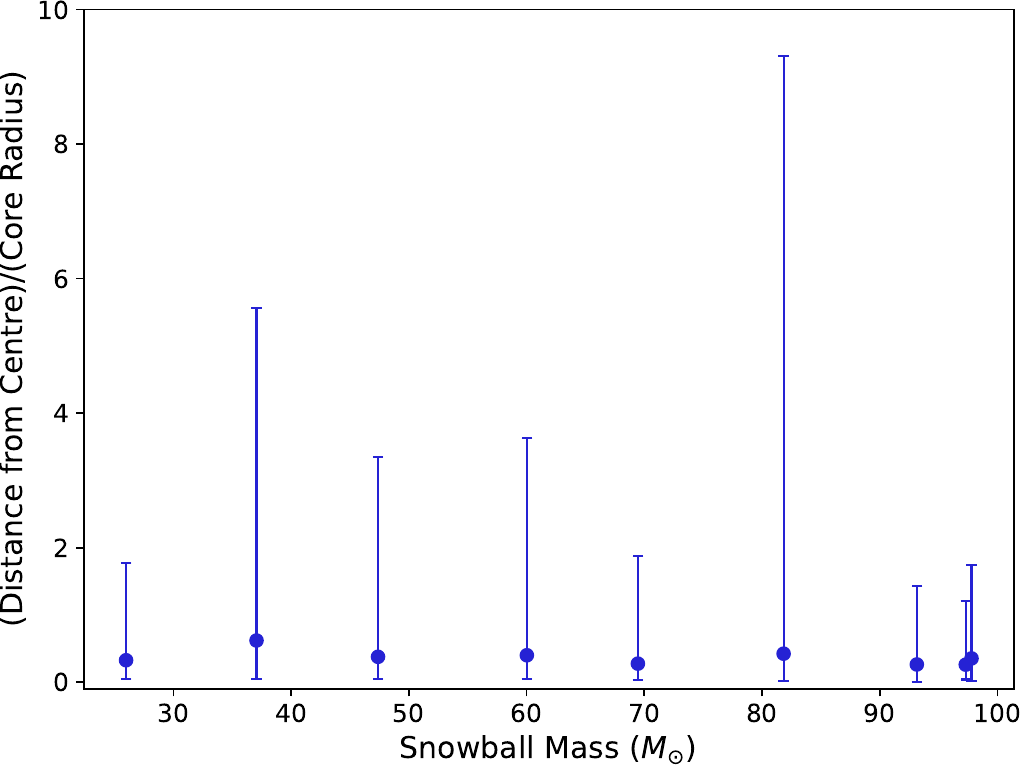}
    \caption{Snowball's position as a function of it's mass. The vertical axis displays Snowball's position, given by a ratio of two distances: Snowball's distance from the cluster centre, and the core core radius. The horizontal axis displays Snowball's mass in solar masses. The blue dots represent the mean position for a given mass, and the error bars displays the range of positions.}
    \label{fig:position}
\end{figure}

\begin{figure*}[htb]
    \centering
    \includegraphics[width=15cm]{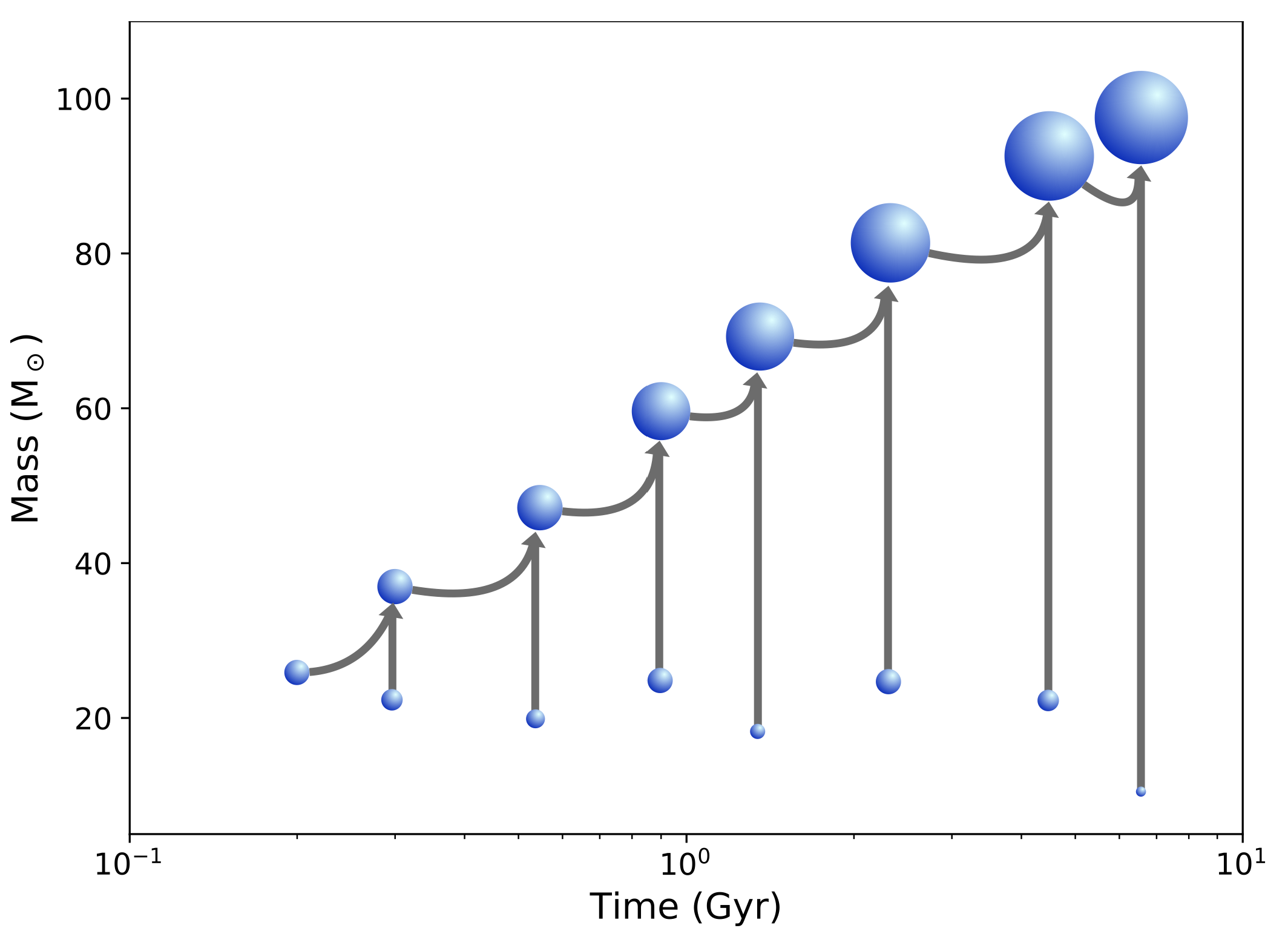}
    \caption{Diagram showing the BH merger tree of Snowball (seven total BHB mergers). Black hole masses are shown on the vertical axis, and the time of mergers is shown on the horizontal axis.} 
    \label{fig:Snowball_mass}
\end{figure*}

\subsection{Snowball's History}

Snowball starts as a 58.1 $\text{M}_{\odot}$ main sequence star. After $\sim 6.4$ Myr the star goes supernova and forms a 26 $\text{M}_{\odot}$ remnant BH, the second most massive BH produced through stellar evolution in the simulation. As this specific simulation does not include primordial stellar binaries, any BHB that emerges must form dynamically through strong interactions. Remarkably, throughout the simulation Snowball is involved in nine mergers; seven BHB mergers, and two involving a main sequence star (when a star-BH merger occurs, \texttt{NBODY6} treats the merger as if the BH fully consumes the star, with no mass loss or merger recoil). All seven BH mergers occur within the cluster core. This is expected, as Snowball spends $95\%$ of its lifetime at $r \lesssim 0.7 r_c $ due to mass segregation resulting from energy equipartition in star clusters over the two-body relaxation timescale (see \citealt{2013MNRAS.435.3272T}). Whilst in the core, Snowball tends to wander around the cluster centre, as has been shown previously of IMBHs within GCs \citep{2018MNRAS.475.1574D}. Figure \ref{fig:position} shows the position of Snowball as a function of its mass. The BH's typical displacement from
the cluster centre appears to be independent of its mass. Snowball’s mass only
changes by a factor of $\sim 4$, and the expected BH displacement depends on ${m_{BH}}^{0.44}$ (see equation
16 in \citet{2018MNRAS.475.1574D}).  Snowball gains mass through each successive BH merger until it reaches its final mass of 97.8 $\text{M}_{\odot}$ as shown in Figure~\ref{fig:Snowball_mass}. All nine mergers occur within the first 6.5 Gyr. Note that all BHB mergers in this simulation occur within this single branch, \emph{i.e.} there are no BHB mergers outside of this merger chain. Hence all Snowball merger companions are first-generation BHs. This is partly due to heavier BHBs being dynamically favourable. Once Snowball becomes the most massive body in the cluster after approximately 230 Myrs, it dominates the dynamics of the core - suppressing the retention of other BHBs.

\begin{figure}
    \centering
    \includegraphics[width = \columnwidth]{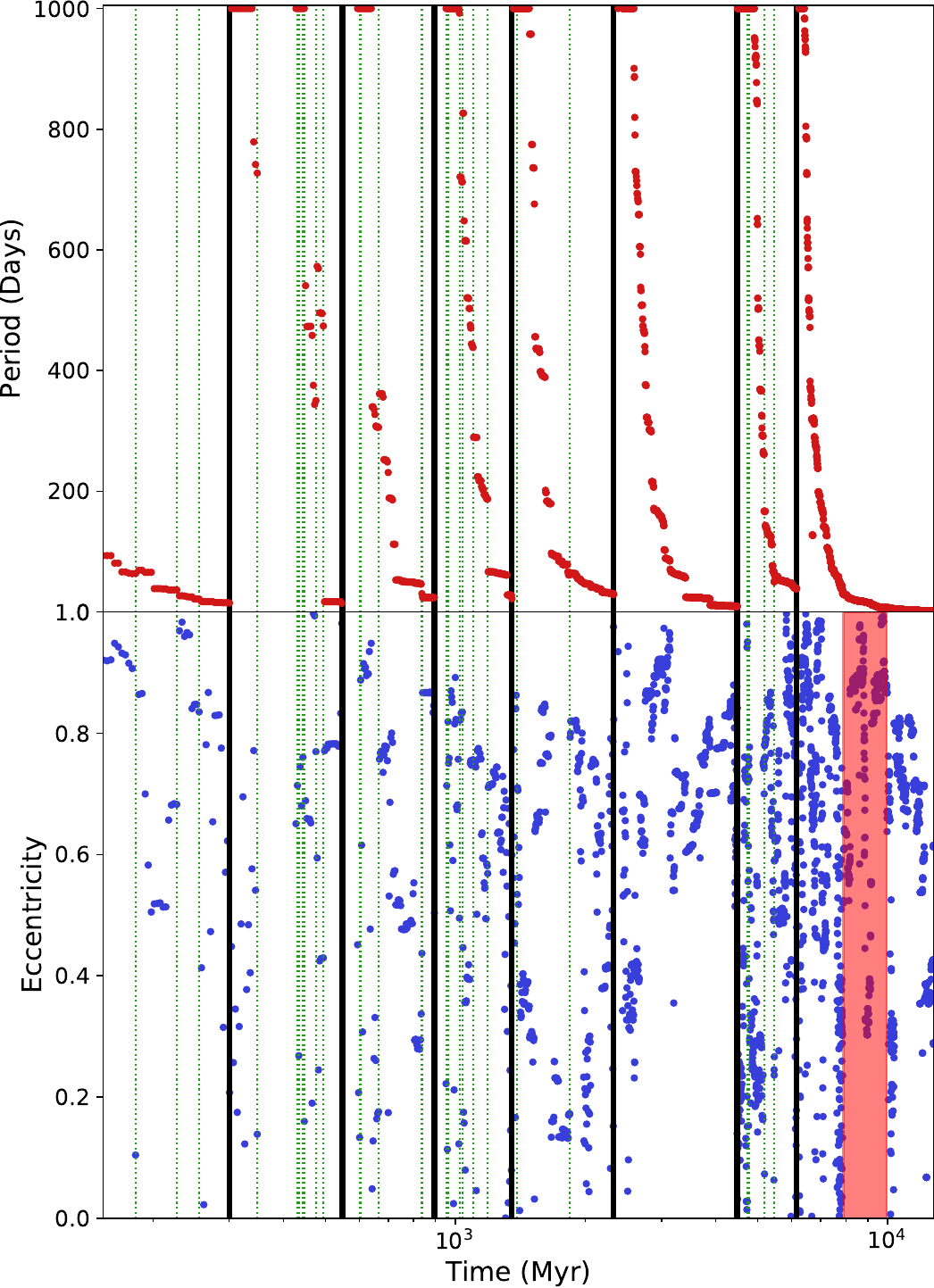}
    \caption{Evolution of the binary period (top panel) and eccentricity (bottom panel) for the various BHBs that Snowball is part of. BHB mergers are displayed as vertical black lines, and three-body exchange events involving the Snowball BHB are displayed as vertical dotted green lines. The eccentricity is given between 0 (circular) and 1 (parabolic), the period in days, and the time in megayears. The time in which Snowball is part of a hierarchical triple BH system is highlighted in red in the bottom panel. Note that there is a sharp upper bound for the periods of 1000 days, as this is the upper cutoff for which \texttt{NBODY6} regularises binaries.}
    \label{fig:eccentricity}
\end{figure}

\medskip \noindent
 Once Snowball migrates to the core by $t \approx 11.7$ Myr, it experiences multiple close interactions with other mass-segregated BHs before forming its first BHB, with an eccentricity of 0.92 and a period of $P = 99$ days. The binary remains in the core for 64 Myr, undergoing various three-body interactions, hardening (with the period reducing to 63 days) and altering the eccentricity before being disrupted, exchanging the secondary component for a 26.14$ \text{M}_{\odot}$ BH, the most massive BH in the cluster at that time. The lower panel in Figure \ref{fig:eccentricity} displays the evolution of Snowball's binary eccentricity. It is clear that the many scattering events significantly alter the BHB eccentricity, with extremes of $e = 0.0008$ (near circular) to $e = 0.999$ (maximally eccentric). After five more three-body interactions the binary is again disrupted at $t \approx 230$ Myr, exchanging the 26.14 $\text{M}_{\odot}$ BH for a 25.34 $\text{M}_{\odot}$ BH and ejecting the heavier former companion from the cluster, making Snowball the most massive body in the cluster. The newly formed binary is fairly eccentric, with $e = 0.97$.

\medskip \noindent
The process of forming a high eccentricity BHB, undergoing multiple strong interactions and an eventual exchange event occurs twice more, until Snowball forms its first merging BHB with a $m_2 = 23.2$  $\text{M}_{\odot}$ companion, corresponding to a mass ratio of 0.89. The binary forms at $t = 258.7$ Myr with $P = 1732$ days and $e = 0.41$. After undergoing nine hardening events the binary separation is lowered sufficiently to allow for significant gravitational radiation, leading to the inspiral and eventual merger at $ t \approx 300$ Myr. This merger nearly doubles Snowball mass to 37.1 $\text{M}_{\odot}$.

\medskip \noindent
Soon after this first merger, at $t = 322$ Myr, Snowball forms a new BHB with a 23.34 $\text{M}_{\odot}$ companion at $r = 0.2 r_{\text{c}}$. The remaining six BHB mergers follow a relatively similar evolution. In the top panel of Figure \ref{fig:eccentricity} we display the evolution of Snowball's binary period. The seven BHB mergers are indicated with vertical black lines. The binaries often form with large periods which rapidly decrease through successive dynamical interactions with other BHs and bodies in the core. We indicate the 26 interactions that cause an exchange by dotted vertical green lines. The rate of BHB hardening tends to decrease as a binary evolves, first hardening rapidly and then slowly reducing the separation until an inspiral occurs, wherein most of a binaries life is spent with small separations. This is partly because the cross-section for three-body interactions of hard binaries is proportional to the semi-major axis \citep{2018arXiv180711489C}. Hence the hardening rate decreases as the binary orbit tightens.

\medskip \noindent
Although most scattering events follow Heggie's law \citep{1975MNRAS.173..729H}, some soften the BHB involved instead. Most of these softening events occur due to an exchange. The attendant interval of chaotic three-body dynamics can dramatically alter the orbital parameters of the newly formed binary, leading to significant softening. This is most evident in the string of exchange events that occur between $ 400 \lesssim t \lesssim  500$ Myr, leading up to the second BHB merger. The top panel of Figure \ref{fig:eccentricity} shows the effects of such interactions, whereby the period varies non-monotonically between 320 and 592 days.

\medskip \noindent 
Scattering events harden binaries and can impart significant recoil velocities which send even IMBHs out of the cluster core \citep{2018MNRAS.475.1574D}. \cite{2020PASA...37...44A} found that $\sim 80\%$ of all ejections follow a three-body encounter with another BH, with the rest being ejected via binary-binary scattering. Although the various BHBs which count Snowball as a member experience significant three-body recoil, the resulting velocities are not sufficient to eject the binary. As can be seen in Figure \ref{fig:position}, at most, the BHB is flung outside the cluster core, where it migrates back towards the centre to undergo further scattering events. One particular interaction leads to a Snowball binary being flung out to $8.45 r_c \approx 1.3$ pc when Snowball is $\sim 82$ M$_{\odot}$. Although Snowball is never ejected, over the $12.5$ Gyrs approximately $67$ binary-single interactions lead to the ejection of a BH. In particular, Snowball plays a large part in the depletion of BHs in the core, being responsible for ejecting $43$ BHs. As the number of BHs in the core is depleted over time (starting from an initial $N_{BH} = 74$ formed from stellar evolution and retained in the cluster after natal kicks) the number of three-body hardening events likewise decreases.

\medskip \noindent
Why is it that no other BHs merge with anything apart from snowball in this cluster?. Throughout the simulation, 10 other BHBs not containing Snowball form in the core, however, none remain bound long enough to either merge or be ejected, with the longest binary only surviving $\sim 31$ Myr. All 10 binaries are disrupted by encounters with Snowball, or with a binary containing Snowball. Seven out of these 10 BHBs undergo an exchange encounter with Snowball, wherein the preexisting binary exchanges a component for Snowball, with the remaining three BHBs simply destroyed with no exchange. As a result of the exchange events, six of the seven newly unbound BHs are ejected from the cluster as a result of the interaction. Snowball's presence in the core is essentially preventing the survival of any other BHB. Exchange encounters are more frequent within the first few gigayears when there is still a sufficiently high number of relatively massive BHs in the core, with only four out of 28 exchanges occurring after $t = 2$ Gyr. This is because exchange events, in general, require BHs of comparable mass \citep{1975MNRAS.173..729H}. On the other hand, scattering events that don't result in an exchange remain relatively common, occurring every few Myr at least until $\sim 10$ Gyr, after which the BH population has mostly evaporated (see also \citealt{2013ApJ...763L..15M} for a Monte Carlo investigation of the BH population in star clusters).

\medskip \noindent
The specifics of Snowball's mass gain depend on the mass of the companion BHs involved in the merger. However, our results clearly show that any BHB within a cluster core is likely to undergo multiple exchange events. Previous scattering experiments also show that the lighter of the two binary components is preferentially ejected from the system in favour of the intruder \citep{1993Natur.364..423S}, meaning that BHBs with mass ratios closer to one are dynamically favoured. This process systematically increases the companion mass before merger. Hence, heavier BHs are more likely to merge than lighter ones, which is indeed what we see here. All seven Snowball BHB mergers involve a secondary BH which is in the heaviest 5\% of all BHs at the time of the merger. This is why most of the BHs that merge with Snowball have masses in the low to mid 20 M$_\odot$, as this is approximately the maximum mass of BHs produced due to our cluster models and supernovae prescriptions.

\subsection{Final State}
By the time Snowball has completed the seventh BHB merger, there are only a total of 19 BHs left in the cluster. Snowball forms another binary at t = 6.65 Gyr, shortly after the last merger. Snowball spends the last 6 Gyr bound to its companion BH at the centre of the cluster, only briefly leaving the core 4 times due to recoil from scattering events (spending a total of 0.032 Gyr out of 6 Gyr at
$r > r_{\text{c}}$).

\medskip \noindent
At $t \approx 7.9$ Gyr, this final binary forms a hierarchical triple system. The system undergoes Kozai-Lidov oscillations \citep{1962P&SS....9..719L,1962AJ.....67..591K}, driving the inner binary to as high as $e = 0.99$ (see the red highlighted section in Figure \ref{fig:eccentricity}). During this time the inner binary still experiences 10 three-body scattering events. Throughout the next 2 Gyr, the outer component of the triple system is exchanged nine times, ejecting the previously bound outer component from the cluster in all but one of these cases. Eventually, the BHB triple system is disrupted at $t \approx 9.94$ Gyr. 

\medskip \noindent
After 10.8 Gyr Snowball and its companion are the only remaining BHs in the cluster. During the last $\sim 200$ Myr, the BHB once again forms a triple system with a white dwarf. This hierarchical triple remains until the end of the simulation.

\section{The Merger Chain With Recoils}
\label{sec:The Merger Chain With Recoils}
The results in Section \ref{sec:snowball} do not include merger recoils caused by gravitational radiation reaction. Such recoils are known to impart kicks comparable to or greater than the escape velocity of the GC \citep{2007ApJ...659L...5C, 2018PhRvL.120o1101R,2018MNRAS.481.2168M, 2020MNRAS.tmp.2087B, Lousto_2010}. Our Monte Carlo method for assigning spin orientations allows us to investigate all possible recoil velocities for each merger so that we can determine how likely it is for the Snowball chain to reach a given length, a probability which is prohibitively expensive to sample adequately with in-run spin assignments. In Section \ref{sec:Preventing The Formation Of Snowball}, we explore statistically how merger recoils impede the formation of a heavy BH like Snowball. In Section \ref{sec:4.2, spin} we discuss the impact of BH spin orientation on the retention or ejection outcome of a merger due to the resulting range of recoil velocities for each of the seven Snowball mergers.

\subsection{Preventing The Formation Of Snowball}
\label{sec:Preventing The Formation Of Snowball}
Including merger recoil velocities prevents the formation of the 97.8 $\text{M}_{\odot}$ BH. The primary issue hindering such a merger chain is that all seven mergers involve a first-generation secondary BH. Hence the mass ratio becomes relatively biased towards low values after the initial merger, leading to significant recoil velocities for subsequent mergers in the chain. Only once the primary BH has grown massive enough to form a BHB with $q \lesssim 0.4$ can recoils once again be low enough for retention, as recoil velocity peaks at $q = 0.4$ (based on Equation \ref{eq:2}). 

\medskip\noindent
BH spins are assigned retroactively in post-processing and are used to calculate BH merger recoils (see section \ref{subsec:Gravitational Recoil}). The two first-generation BHs involved in the initial merger in the Snowball chain are assigned spin magnitudes of $\alpha_1 = 0.08$ and $\alpha_2 = 0.18$, based on the \texttt{MESA} code. From the merger between these two BHs, the resulting second-generation BH is ejected from the cluster in approximately $85.2\%$ of the $10^4$ merger recoil calculations. In approximately $14.5\%$ of the $10^4$ recoil calculations the second generation BH merges again, and the resulting third generation BH is ejected. In approximately $0.430\%$ of the $10^4$ recoil calculations the third generation BH merges and the resulting fourth generation BH is ejected. The longest Snowball chain, comprising four mergers, occurs twice in $10^4$ recoil calculations.

\subsection{Spin}
\label{sec:4.2, spin}
\begin{figure}
    \centering
    \includegraphics[width = \columnwidth]{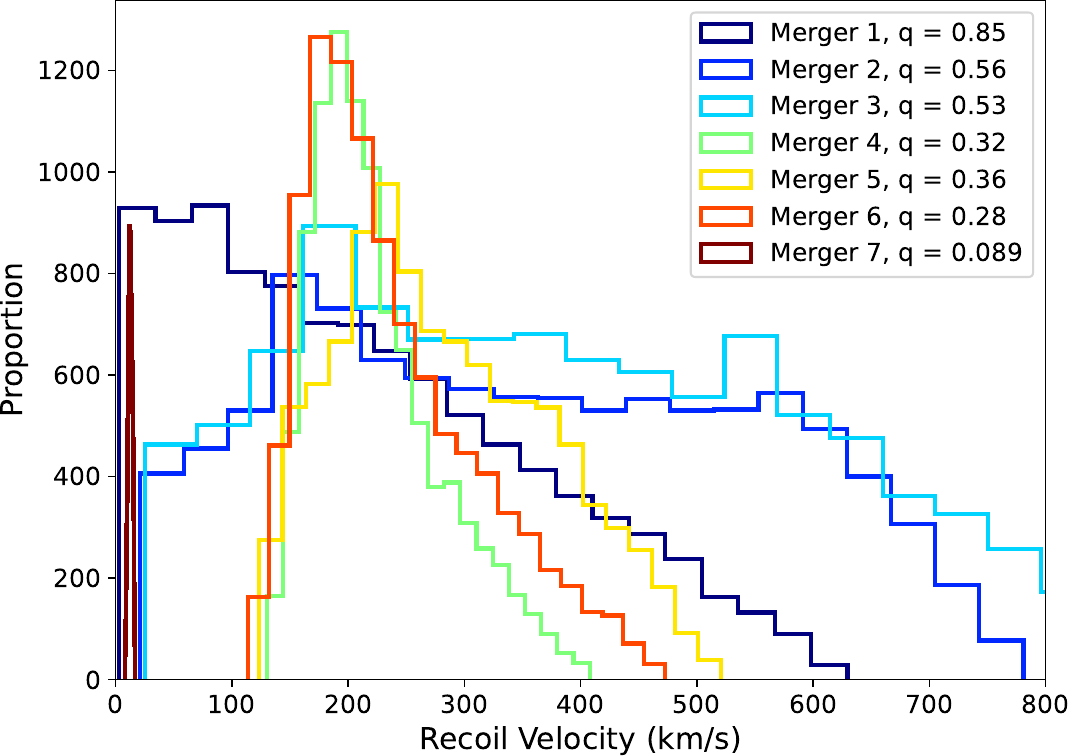}
    \caption{Histogram of the various possible recoil velocities for each of the 7 mergers in the Snowball merger chain. For each merger we calculate 10,000 different recoils, randomly assigning the spin directions isotropically, but keeping all other properties such as mass, initial separation and spin magnitudes fixed. Mergers are labelled 1 - 7 in the order in which they occur, along with the corresponding BHB mass ratio.}
    \label{fig:recoils}
\end{figure}

The above results highlight the strong dependence of spin direction on merger recoil velocity. As discussed in sec \ref{subsec:Gravitational Recoil}, we calculate $10^4$ different recoils for each \texttt{NBODY6} merger, fixing the mass ratio and spin magnitude but randomly assigning spin directions. As a pedagogical device, Figure \ref{fig:recoils} shows the histogram of possible recoil velocities for each of the seven mergers in the original, zero-recoil Snowball chain, ignoring temporarily that recoils truncate the chain after a maximum of four mergers. As the mass ratio decreases with subsequent mergers, larger recoils become more likely, until $q$ drops below 0.4, as discussed above. The BHs produced through mergers also tend to have significantly higher spin magnitudes than those formed through core collapse, as the majority of the BHB orbital angular momentum is transferred to the spin of the merger product. Higher spin magnitudes increase the impact of spin directions on the final velocity. From Figure \ref{fig:recoils} we see why the Snowball chain never extends past four mergers, as the minimum recoil velocity of the fourth merger is $\sim 129 \text{kms}^{-1}$ (green line in Figure \ref{fig:recoils}), far above the host cluster's central escape velocity. The recoil distribution for the seventh merger (brown line in Figure \ref{fig:recoils}) displays a uniquely narrow range of possible velocities due to the extreme mass ratio of $q = 0.089$. The histogram peaks at $< 10 \text{kms}^{-1}$, with a median of $6.14 \text{kms}^{-1}$ and a standard deviation of $4.83 \text{kms}^{-1}$. However, due to the larger recoils from earlier mergers in the chain, Snowball would be ejected long before it could undergo this seventh merger. 


\medskip \noindent
Although gravitational recoils eject Snowball after at most only four mergers, longer merger chains can occur rarely. One could foresee a scenario whereby an initial merger forms a BH that is not only retained but is also substantially larger than the remaining BH population. This massive second-generation BH could then merge multiple times with relatively light BHs without being ejected, as the low mass ratios would lead to low recoil velocities.

\medskip \noindent

\section{Merger Chain Statistics}
\label{sec:rates}
 Although the Snowball chain only reaches seven mergers if we ignore the effects of gravitational recoil, it is still instructive to discuss statistics of such a chain under idealised conditions, as a baseline benchmark against which simulations with recoil can be compared, as the understanding of recoil physics develops in the future. Generally, the no-spin case represents the optimal path to form a massive BH via repeated mergers for a system with a shallow potential like GCs, hence this scenario gives an upper limit to the probability of finding an IMBH through repeated mergers. Current models for gravitational recoils are only approximate as they rely on fitting and/or extrapolation to provide a general analytical solution to an intrinsically complex numerical relativity problem \citep{Lousto_2010, Zlochower_2015, 2010ApJ...719.1427V, 2007ApJ...659L...5C}. By comparing results from this section with future GW detections of massive BHs, it will be possible not only to constrain dynamical formation pathways, but also falsify theoretical parametrizations of BH recoil in a specific astrophysical context.

\medskip \noindent
To estimate the likelihood of encountering Snowball-like chains in typical compact star clusters, we consider our full set of $N=58$ simulations and construct a probability density function of merger chain length. We define the merger chain length $L$ as the total number of BHB mergers involved in producing the final multi-generation BH. Based on this definition, Snowball has a chain length of seven.
We also denote the generation of a BH as $\text{BH}_{\text{gen}}$. Across all 58 simulations, we observe 18 chains with $L=1$, 28 chains with $2 \leqslant L \leqslant 4$, and Snowball with $L=7$.

\subsection{Chains With And Without Recoil}
\label{subsec:Chains With And Without Recoil}

\begin{figure}
    \centering
    \includegraphics[width = \columnwidth]{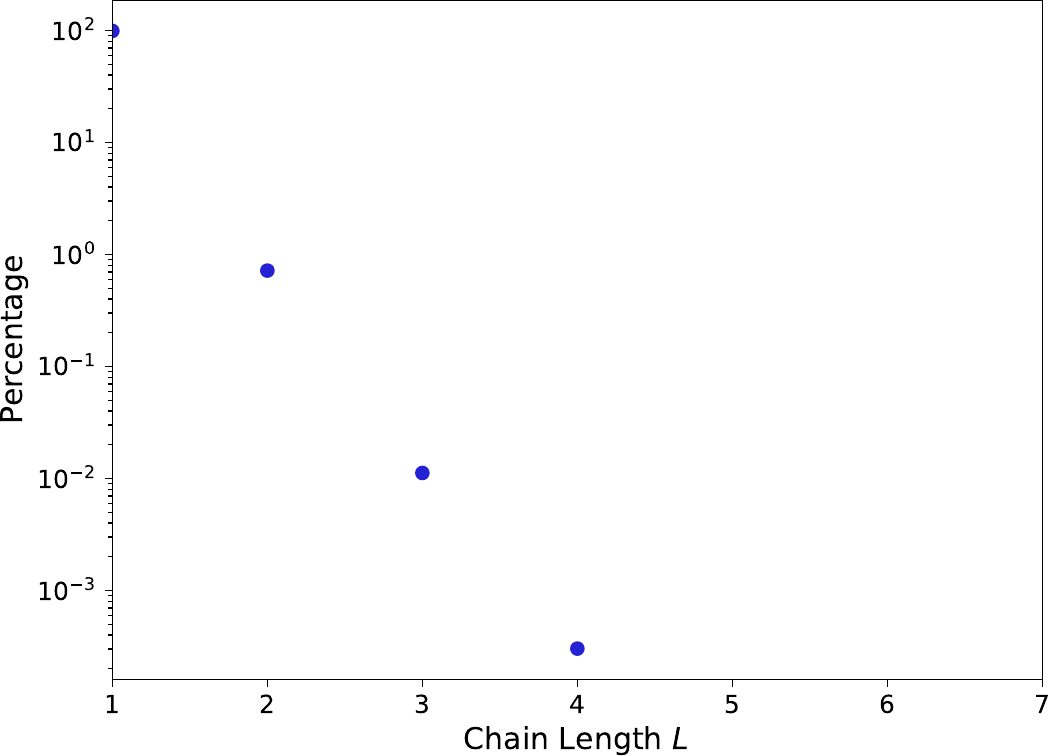}
    \caption{The percentage of chains with length $L$ from our Monte Carlo method used to calculate recoil velocities with randomized spin in post-processing, plotted on a vertical log scale. Approximately $99.3\%$ of the chains reach $L=1$, approximately $0.719\%$ reach $L=2$, approximately $0.0112\%$ reach $L=3$, and approximately $3.06\times10^{-4}\%$ reach $L=4$. There are no chains with length $L \geqslant 5$.}
    \label{fig:recoil_percentages}
\end{figure}

Our post-processing implementation of gravitational recoil allows us to directly compare the merger chains statistic for recoil enabled and disabled. Table \ref{tab:BH_generation_table} lists the number of merger chains both with and without gravitational merger recoil taken into account in post-processing. Without recoil, there are a total of 97 mergers. We highlight the presence of four chains of length $L=4$, the absence of chains with $5\leqslant L \leqslant 6$, and the existence of Snowball with $L=7$. Moreover, only three chains involve two BHs with $\text{BH}_{\text{gen}} > 1$, because BHs with BH$_{\rm gen} \geqslant 2$ constitute a small fraction of the BH population. The number of $L>1$ chains is greatly reduced when including recoil, although without spin there are still three $L=3$ and $L=4$ chains. There are only 70 BH mergers with recoil (no-spin) included. To account for the effects of spin orientation on merger recoil, we simulate each of the 47 merger chains 10,000 times (see section \ref{subsec:Gravitational Recoil}). Figure \ref{fig:recoil_percentages} displays $P(L)$, the percentage of chains with length $L$ from our Monte Carlo method. We find that approximately $99.3\%$ of the chains reach $L=1$, approximately $0.719\%$ reach $L=2$, approximately $0.0112\%$ reach $L=3$, and approximately $3.06\times10^{-4}\%$ reach $L=4$. We find no $L \geqslant 5$ chains. The third column in Table \ref{tab:BH_generation_table} displays the number of chains if the most probable recoil velocity for each merger is chosen.

\medskip \noindent
This is expected, as longer chains are less likely regardless of gravitational recoil, as longer chains are multiplicative in nature. At a most basic level, we expect the probability of a certain chain length to follow

\begin{equation} \label{eq:power_law}
    P(L) \propto P_o^{L},
\end{equation}

\noindent

where $P_o$ is the probability of getting any $L > 1$ chain. Notably, the functional form of equation \ref{eq:power_law} is consistent with Figure \ref{fig:recoil_percentages}.
This explains the rarity of longer merger chains. Fitting to the zero-recoil case, we estimate a probability of a Snowball merger to be $P(L=7) \approx 0.022$, consistent with observing its occurrence once out of 47 chains. If we instead fit for the recoil case without spin, we can loosely extrapolate out to $L=7$, finding $P(L=7) \approx 1.4\times 10^{-4}$. For the case with spin, we can extrapolate from our results in Figure \ref{fig:recoil_percentages}, finding $P(L=7) \approx 2.5\times 10^{-12}$.

\begin{table} [htb]
   
	\begin{center}
	 \caption{Prevalence of different merger chain lengths}
	 \label{tab:BH_generation_table}
	\begin{tabular}{cccc} 
	
		\hline
		& \multicolumn{3}{c}{\textbf{Number of chains}}\\
		\hline
		\textbf{L} & \textbf{No recoil} & \textbf{Recoil (no-spin)} & \textbf{Recoil (spin)} \\
		\hline
		1 & 18 & 33 & 41 \\
		2 & 16 & 8  &  7 \\   
		3 & 8  & 3  & 1 \\  
		4 & 4  & 3  & 0 \\
		5 & 0  & 0  & 0 \\
		6 & 0  & 0  & 0 \\
		7 & 1  & 0  & 0 \\
		\hline
		\textbf{Total} & 47 & 47 & 49  \\
		\hline
	\end{tabular}
	\end{center}
	\tablecomments{ We report the number of chains both with and without gravitational recoil taken into account. Note that the numbers listed for the Recoil (spin) case are when we simply select the most likely recoil velocity for each merger. The total number of merger chains increases to 49 when including spin. The increase is a result of the two $L=4$ chains which involve the merger between $\text{BH}_{\text{gen}} = 2$ and $\text{BH}_{\text{gen}} = 3$ BH (which are present in the no recoil and recoil (no-spin) cases) being cut short, as the $\text{BH}_{\text{gen}} = 3$ object is ejected due to the high merger recoil. This leads to two extra $L=2$ chains. The merger chains correspond to 97, 70 and 58 BHB mergers for no recoil, recoil, and spin recoil respectively.}
	
\end{table}

\medskip \noindent
Without recoil, approximately half of the 97 in-cluster mergers only involve BHs produced directly through core collapse. 18 of the resulting second-generation BHs never merge again, increasing to 33 when merger recoils are accounted for (without BH spin). 47 mergers involve a BH with $\text{BH}_{\text{gen}} \geqslant 2$. In 36 of these mergers $m_2$ is a first-generation BH, with $m_1$ being the multi-generational object. This is expected, as hierarchically formed BHs are on average heavier than those formed through core collapse, and thus are more likely to be the heavier element in a BHB. Mergers involving an object with BH$_{\text{gen}} \geqslant 2$ involve a first-generation BH merging with the higher generation BH, with only three exceptions. One exception is a merger between two second-generation BHs, and the other two are between a second and third-generation BH. These three mergers represent the only such mergers between two previous merger remnants. When we take into account BH spins, the only merger between two multi-generational BHs is the merger between the two second-generation BHs. Hence we conclude two or more chains are unlikely to exist simultaneously within a mid-sized GC ($\sim 10^5$ stars). This is expected, as once a BHB merges, its product is likely to be the most massive body in the cluster, and thus sink to the cluster centre due to dynamical friction, even if merger recoil is imparted (as long as the resulting BH remains bound). The massive BH dominates dynamically within the dense core, likely undergoing many dynamical interactions with other single and binary BHs. Hence, any BHB that forms is likely to interact with the second-generation BH, in which case an exchange event will preferentially unbind $m_2$, forming a new BHB with the second-generation BH \citep{1993Natur.364..423S}. It is important to note that recoil velocity is approximately independent of the total binary mass, but does depend on $q$. Hence light and heavy mergers have the same chance of reaching a given escape velocity.

\medskip \noindent
  Including recoil significantly increases the number of post-merger BH ejections, resulting in most larger merger chains no longer being viable. The $L=7$ chain no longer occurs, now only reaching a chain length of four at most before the BH is ejected. Even without spin, the recoil is often large enough to eject the majority of the 50 second-generation BHs, with only 23 being retained, 17 of which undergo further mergers. Only five of the 13 third-generation BHs are retained, all of which merge again. This suggests that any retained BHs with BH$_{\text{gen}} > 1$ have a high chance of merging again. Including spins in the recoil calculations tends to increase the size of the kicks, with now only approximately $0.731\%$ of the second-generation BHs retained after formation, $\approx 73.9\%$ of which merge again. Approximately $85\%$ of the resulting $L = 3$ chains correspond to a merger between two BH$_{\text{gen}} = 2$ objects, forming a $58 \text{M}_{\odot}$ BH. This is likely the main way an $L = 3$ chain can occur in GCs of these sizes, as the other pathway involves a BH $_{\text{gen}} = 1$ object merging with a BH$_{\text{gen}} = 3$ object (as in the Snowball merger chain). This pathway is less likely as it requires the retention of a BH$_{\text{gen}} = 3$ object, which only occurs approximately $0.450\%$ of the time in the case of Snowball's merger chain. If we take the most likely recoil for each merger based on all spin orientations, we find only 58 BHB mergers, meaning that only $3.3\%$ of all the BHs that remain bound after natal kicks end up merging.
  
 
\subsection{Snowball event Rate Estimation}

The results in Section \ref{subsec:Chains With And Without Recoil} can be converted into a rough estimate of the IMBH formation rate in GCs. Firstly we consider the no-recoil event rate as a baseline estimate. To estimate the comoving rate of Snowball-like merger chains to an order of magnitude, we can use the equation


\begin{equation}\label{eq:OoMrate}
    R_{snowball} \approx \frac{R_* f_{\text{GC}} f_{\text{Snowball}}}{M_{GC}},
\end{equation}

\noindent \medskip
where $R_*$ is the comoving star formation rate [units: $M_\odot \text{yr}^{-1}\text{Mpc}^{-3}$] at the characteristic time of star cluster formation and within the volume observed by LIGO, $f_{\text{GC}} \approx 10^{-3}$ is an order-of-magnitude estimate of the fraction of star formation that occurs within a dense star cluster, $f_{\text{Snowball}}$ is the fraction of those clusters that are likely to produce Snowball-like merger chains, estimated from our simulations, and $M_{GC} \approx 6.5\times10^4 M_\odot$ is the characteristic mass of a dense star cluster in our runs.

\medskip
\noindent 
A detailed estimation of $R_{\text{snowball}}$ would require to take into account both the redshift dependence of $R_*(z)$ and the distribution of time delays between the formation of the cluster and a chain merger event. However, uncertainties in $f_{\text{GC}}$, $f_{\text{Snowball}}$, and $M_{\text{GC}}$ dominate Eq.~\ref{eq:OoMrate}, thus to remain in the spirit of an estimate to only an order of magnitude, we can neglect redshift dependency for $R_{*}$, adopting an appropriate average value for it, and assume the merger time delay distribution is a delta function at the cluster's age. Given that  LIGO is sensitive to high mass merger events out to $z\sim 1 - 1.5$ \citep{2021CQGra..38e5010C}, and that most of the volume is thus at cosmological distances, we assume $R_* \approx 0.1 M_\odot \text{yr}^{-1}\text{Mpc}^{-3}$, corresponding to the comoving star formation rate at $z \approx 1.2$ from \citet[Figure 9, left panel]{2014ARA&A..52..415M}. We also neglect the effects of a cluster birth mass function term. For our basic estimate, we simply ignore clusters with stellar masses smaller than our simulations, and use characteristic values for the population of globular clusters in the Milky Way. To an order of magnitude, a more detailed calculation would give the same result as although $f_{\text{Snowball}}$ would be smaller in lower mass clusters, one would have to increase $f_{GC}$ to reflect the fact that a greater proportion of star formation is considered in the calculation if lower mass clusters are included.


\medskip
\noindent 
Based on the results presented in Table \ref{tab:BH_generation_table}, we assume $f_{\text{Snowball}} \approx 0.02$ (one IMBH chain event out of over 50 simulations). A more precise characterisation is challenging from a single chain instance discovered in our set of simulations. Given these assumptions, we estimate the Snowball event rate to be

\begin{equation}
    R_{\text{snowball}} \approx 0.03   \text{Gpc}^{-3} \text{yr}^{-1}.
\end{equation}

\noindent
This is within the 90\% confidence interval of the inferred rate of mergers similar to GW190521, estimated by \citealt{Abbott_2020} as $0.13\pm^{0.3}_{0.11}\text{Gpc}^{-3} \text{yr}^{-1}$. We can take $f_{\text{Snowball}} \approx 0.02$ as an estimate for the upper limit for our cluster models. The fact that this upper limit does not lead to an event rate exceeding the observed LIGO rate is reassuring to qualitatively support the plausibility of IMBH formation through chain mergers in dense stellar systems, but of course, a deeper quantitative investigation is needed to further assess the scenario we propose.

\medskip
\noindent 
Of course, the above agreement is only for the special baseline case without merger recoil. Assuming the rough estimate for a Snowball chain rate of $P(L=7) \approx 10^{-12}$ with spin recoil, presented in Section \ref{subsec:Chains With And Without Recoil}, we obtain $f_{\text{Snowball}} \approx 0.02/(8.8\times 10^{9}) = 2.3\times 10^{-12}$, and hence

\begin{equation} \label{eq:_recoil_rate}
    R_{\text{snowball}} \approx 3.5\times10^{-12}   \text{Gpc}^{-3} \text{yr}^{-1}.
\end{equation}


\noindent
Although this event rate is orders of magnitudes bellow even the lower bound of the inferred LIGO rate, an under-prediction is expected, as there are other potential GW190521 formation channels. Our estimates are also based on our cluster models, and one would expect $f_{\text{Snowball}}$ to be larger for more massive clusters with larger escape velocities. Note that Equation \ref{eq:_recoil_rate} is simply based on Equation \ref{eq:power_law}, extrapolated out to $L=7$. 

\section{Discussion} 
\label{sec:implications}

In this work we analyze the hierarchical mergers of BHBs formed within simulated GCs, investigating the dynamical formation of an IMBH with and without merger recoils. The analysis is based on a set of direct \texttt{NBODY6} simulations of realistic median sized cluster models previously presented in \cite{2020PASA...37...44A}. We present a detailed investigation into the evolution of ``Snowball", a BH that undergoes seven BHB mergers in a baseline system excluding merger recoil, attaining a final mass of $\sim 98$ $M_\odot$. We also explore the effects of gravitational merger recoil on this  Snowball merger chain, and on the prevalence of different length merger chains across our 58 simulations. Through our Monte Carlo method of calculating merger recoils, we estimate the probabilities of merger chains reaching a given length.
 
\medskip \noindent
BH spins can lead to recoil velocities above $10^3 \text{kms}^{-1}$ \citep{2007ApJ...659L...5C}, well above the central escape velocity of any known GC. Hence merger chains with $L > 2$ are unlikely, as backed up by our results. Based on Table \ref{tab:BH_generation_table}, we predict that median mass GCs are more likely to host a single $L>1$ merger chain than multiple chains. We therefore conclude that long chains of seven or more mergers are too rare to explain the full rate inferred by LIGO data, without substantial modifications to the currently accepted fitting formulas for BH recoil, or some other modification to the simulation physics unrelated to recoil. However, we do find a small percentage of $L=3$ and $L=4$ chains,  indicating that even with the current analytical formulation of BH recoils, there is a region of the parameter space of initial conditions (BH masses, spins) where the resulting recoil is insufficient to eject the merger product. The majority of the $L=3$ chains involve the merger between two second-generation BHs. If we use our estimated probability for an $L=3$ chain of $P(L=3) = 4.95\times 10^{-5}$, we estimate $R_{\text{L=3}}  \approx 1.5 \times 10^{-4}  \text{Gpc}^{-3} \text{yr}^{-1}$. This is a potential formation pathway for GW190521. Although due to our initial conditions and supernovae prescriptions the second-generation BHs in these $L=3$ chains are lighter than the primary and secondary masses of GW190521, a similar chain with heavier initial first-generation BHs could potentially explain the heavy merging masses observed by LIGO/Virgo. In fact, through the analytic formulae for gravitational recoil presented in Section \ref{subsec:Gravitational Recoil} (eq \ref{eq:1}-\ref{eq:4}), and merger mass loss \citep{Lousto_2010}, we find that that GW190521 could have formed in a median mass GC from an $L = 3$ merger chain. In this scenario, both the primary and secondary components are second-generation BHs that have been retained post-merger due to sufficiently low recoil velocities (like the $L = 3$ chain we observed in our simulations), whilst still achieving the required mass observed by LIGO.

\medskip \noindent
The scenario above requires fine-tuning of a specific set of mass and spin combinations for the precursor first-generation BH. We can calculate the recoil and mass losses for all realistic BH mass-ratio and spin combinations, and find the proportion of combinations that lead to a retained post-merger BH matching the masses of the GW190521 primary and secondary BHs. Not all combinations are viable, as low mass ratio binaries are dynamically unfavourable, due to being more likely to be disrupted via exchange events. To account for this we restrict our calculation to mass ratios above 0.4, as this is the lowest mass ratio of any first-generation merger across all of our simulations. We also restrict the spin magnitudes to those outputted from the MESA fitting formula (see section \ref{subsec:Gravitational Recoil}). With these restrictions in place, we assume that every mass-ratio and spin combination is equally likely. Spin directions are selected isotropically. We also assume that the upper mass gap cuts of at 50 $\text{M}_\odot$, we find that for a cluster escape velocity of 25 $\text{ms}^{-1}$, approximately 0.29\% of the possible combinations lead to retention of a sufficiently massive primary ($\sim 85 \text{M}_\odot$), and 0.38\% of combinations for a secondary ($\sim 66 \text{M}_\odot$). Although the chance of both mergers occurring in the same cluster and then merging again in a GW190521 like event is rare, it is still a viable formation scenario. The set of viable combinations is increased if the cluster has a higher escape velocity, which likewise increases the likelihood of such a merger chain. If we instead set the escape velocity at 50 $\text{ms}^{-1}$, then the proportion of viable mass and spin combinations increases to 1.5\% and 1.7\% for the primary and secondary masses respectively. This may suggest that GW190521 originated in a star cluster more massive than those we simulated. 


\medskip \noindent
On average across our 58 simulations, the last merger in chains with $L > 1$ occurs at $t = 4735$ Myr. Thus it appears that hierarchical mergers in relatively young GCs ($t \lesssim 6$ Gyr) are preferred, when there is still a sufficient population of BHs in the core. If true, this means that hierarchical mergers are more likely to occur at cosmological distances, when the star formation rate is higher and there was a higher proportion of young GCs. In principle, such distant sources are more difficult to detect. However, the larger average chirp mass from hierarchical mergers leads to a larger gravitational strain.



\medskip \noindent
The hierarchical merger pathway provides a natural method for forming BHs within the BH mass gap, without the need for new physics, but that does not mean that such new physics is absent. Although gravitational recoil lowers the likelihood of longer merger chains occurring, the formation of multi-generation BHs, including those similar to the GW190521 final BH is still possible within mid-sized GCs. Regardless of the origin of GW190521, this remarkable detection represents the first confirmed detection of an IMBH and will likely spark increased interest in the formation and evolution of these stellar behemoths. 


\acknowledgments

We would like to thank Morgan Macleod for useful discussion. Parts of this research are supported by the Australian Research Council Centre of Excellence for Gravitational Wave Discovery (OzGrav) (project number CE170100004). Numerical simulations have been performed on HPC clusters at the University of Melbourne (Spartan) and Swinburne University (OzSTAR).




\bibliography{bibliography}{}
\bibliographystyle{aasjournal}

\end{document}